\documentclass[sigconf]{acmart}
\AtBeginDocument{%
  \providecommand\BibTeX{{%
    \normalfont B\kern-0.5em{\scshape i\kern-0.25em b}\kern-0.8em\TeX}}}

\setcopyright{acmlicensed}
\copyrightyear{2024}
\acmYear{2024}
\acmDOI{XXXXXXX.XXXXXXX}

\acmConference[DAC '24]{61st ACM/IEEE Design Automation Conference}{June 23--27, 2024}{San Francisco, CA}
\acmISBN{978-1-4503-XXXX-X/18/06}

\begin{document}

\title{Invited: Neuromorphic architectures based on augmented silicon photonics platforms}


\author{Matěj Hejda}
\orcid{0000-0003-4493-9426}
\affiliation{%
  \institution{Hewlett Packard Labs, Hewlett Packard Enterprise}
  \streetaddress{Hermeslaan 1A}
  \city{Diegem}
  \country{Belgium}}
\email{matej.hejda@hpe.com}

\author{Federico Marchesin}
\affiliation{%
  \institution{Ghent University - imec}
  \city{Ghent}
  \country{Belgium}
}
\email{federico.marchesin@ugent.be}

\author{George Papadimitriou}
\affiliation{%
  \institution{Department of Informatics and Telecomm., University of Athens}
  \streetaddress{30 Shuangqing Rd}
  \city{Athens}
 \country{Greece}}
 \email{georgepap@di.uoa.gr}

\author{Dimitris Gizopoulos}
\affiliation{%
  \institution{Department of Informatics and Telecomm., University of Athens}
  \streetaddress{30 Shuangqing Rd}
  \city{Athens}
  \country{Greece}}
  \email{dgizop@di.uoa.gr}

\author{Benoit Charbonnier}
\affiliation{%
  \institution{Univ. Grenoble Alpes, CEA, LETI}
  \streetaddress{8600 Datapoint Drive}
  \city{Grenoble}
  \country{France}}
\email{benoit.charbonnier@cea.fr}

\author{Régis Orobtchouk}
\affiliation{%
  \institution{Univ. Lyon, ECL, INSA-Lyon, UCBL, CPE Lyon, CNRS, INL}
  \streetaddress{}
  \city{Lyon}
  \country{France}}
\email{regis.orobtchouk@insa-lyon.fr}

\author{Peter Bienstman}
\affiliation{%
  \institution{Ghent University - imec}
  \streetaddress{1 Th{\o}rv{\"a}ld Circle}
  \city{Ghent}
  \country{Belgium}}
\email{peter.bienstman@ugent.be}

\author{Thomas Van Vaerenbergh}
\affiliation{%
  \institution{Hewlett Packard Labs, Hewlett Packard Enterprise}
  \streetaddress{Hermeslaan 1A}
  \city{Diegem}
  \country{Belgium}}
\email{thomas.van-vaerenbergh@hpe.com}

\author{Fabio Pavanello}
\orcid{0000-0002-6889-2012}
\affiliation{%
  \institution{Univ. Grenoble Alpes, Univ. Savoie Mont Blanc, CNRS, INPG, CROMA}
  \streetaddress{3 parvis Louis Néel, CS 50257}
  \city{Grenoble}
  \country{France}
  \postcode{38016}
}\email{fabio.pavanello@cnrs.fr}

\renewcommand{\shortauthors}{Hejda, Marchesin et al.}

\begin{abstract}
In this work, we discuss our vision for neuromorphic accelerators based on integrated photonics within the framework of the Horizon Europe NEUROPULS project. Augmented integrated photonic architectures that leverage phase-change and III-V materials for optical computing will be presented. A CMOS-compatible platform will be discussed that integrates these materials to fabricate photonic neuromorphic architectures, along with a gem5-based simulation platform to model accelerator operation once it is interfaced with a RISC-V processor. This simulation platform enables accurate system-level accelerator modeling and benchmarking in terms of key metrics such as speed, energy consumption, and footprint.
\end{abstract}







\maketitle

\section{Introduction}
Recent advances in deep learning (and more recently genAI) empower machines with remarkable information processing and synthesis capabilities. At the same time, the exponential growth in data volumes generated by an extensive range of consumer devices and industrial sensors calls for novel approaches to advanced and efficient data processing. Although advanced AI models are typically deployed in purpose-built high-performance cloud clusters, some applications require advanced data processing locally at the edge, that is, closer to where data were originally generated \cite{cao_overview_2020}. More specifically, edge computing requires the development of lightweight accelerators capable of providing AI-tailored data processing locally with low latency and high energy efficiency \cite{satyanarayanan_emergence_2017}. 
\begin{figure*}[t]
  \centering
  \includegraphics[width=0.8\textwidth]{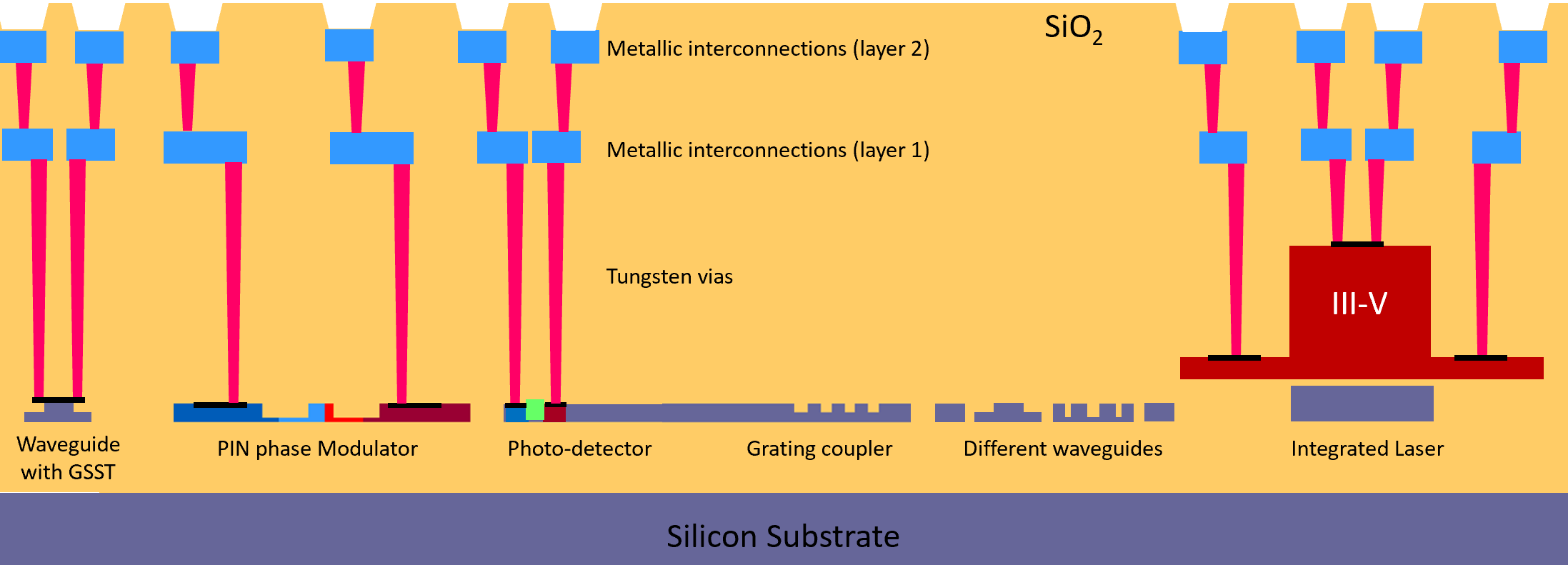}
  \caption{CMOS-compatible augmented platform in the NEUROPULS project. Reprinted with permissions from \cite{pavanello_neuropuls_2023}. IEEE\copyright}
  \label{fig:platform}
\end{figure*}
Current state-of-the-art AI is enabled by graphical processing units (GPUs), tensor processing cores and similar application-specific hardware that allows for high computing parallelization and optimized computation of linear algebraic operations that underpin modern deep learning workloads \cite{samek_explaining_2021}. In line with other conventional computing approaches, these digital architectures utilize Von Neumann processor architectures with distinct memory and computing blocks. This poses some inherent limitations in terms of high demands on data movement between these functional blocks in terms of bandwidth, energy efficiency, and power consumption.

The goal of alleviating these bottlenecks is one of the main objectives in fields such as in-memory computing, where (a part of) the data needed for the computation is co-located with the computing core architecture, thus requiring less data movement \cite{zhou_-memory_2023}. In neuromorphic (brain-inspired) engineering, the principle of memory-compute co-location is further enhanced with signaling and information processing principles analogous to those observed in the brain, which promises further improvements in computation sparsity and energy efficiency \cite{Frenkel2023_PotI}. Although significant efforts have been devoted to leveraging existing CMOS technology to develop digital and mixed-signal neuromorphic computing cores \cite{hoppner_spinnaker_2022,davies_advancing_2021}, alternative technologies such as spintronics or photonics are nowadays gaining considerable momentum for the implementation of next-generation computing hardware \cite{grollier_neuromorphic_2020,Shastri2021_NatPhot}. However, there are still certain practical limitations towards achieving accelerators based on integrated photonics. Current CMOS-compatible silicon photonic platforms do not provide a full range of required functional modalities. In addition, photonic accelerators are often investigated standalone, rather than being interfaced with a system-level architecture and used in real-world scenarios. Finally, full-scale, system-level simulation platforms of a photonic accelerator interfaced with a processor core remain mostly unexplored. In the next sections, we will discuss how we are aiming to tackle all these aspects within the framework of the NEUROPULS project.

\section{CMOS-compatible SiPh platform}
 
Unlike in the electronic neuromorphic approaches, photonics allow one to leverage desirable properties of lightwaves such as wavelength multiplexing, low-loss signal propagation without Joule heating as well as access to very high bandwidth devices (above tens of GHz) to e.g. encode or read-out data \cite{tait_silicon_2019}. In particular, integrated photonics has emerged as a promising size, weight and power (SWaP)-optimized platform. Silicon photonics (SiPh) currently represents arguably the most promising approach to photonic integration due to its compatibility with existing CMOS approaches for cost and volume. However, certain functionalities are not available in pure Silicon-On-Insulator (SOI) platforms. In particular, the ability to realize non-volatile optical memory elements is one of such missing functionalities that is key to achieving energy-efficient optical computing architectures \cite{feldmann_all-optical_2019,feldmann_parallel_2021}. In addition, active devices (such as lasers) cannot be realized in silicon because of its indirect bandgap. Therefore, additional materials such as III-V compound semiconductors need to be co-integrated into SOI platforms, typically using heterogeneous or hybrid integration methods. A monolithic fabrication approach is the most desirable way to achieve these missing functionalities in a compact manner with excellent alignment tolerances between patterned layers and consequent ease of coupling, as well as packaging costs lower than those of hybrid approaches. Such platform has not yet been developed or presented as an open access service, as is the case for pure SOI platforms \cite{rahim_open-access_2018}. 

One of the goals of the Horizon Europe NEUROPULS project is to develop a platform that can accommodate these additional functionalities in a monolithic manner. In Figure~\ref{fig:platform}, the integration strategies for PCMs (e.g., GSST or other types) and III-V materials are shown. This integration does not affect other building blocks, such as high-speed modulators and detectors (above 50 GHz bandwidth) that have already been developed for the platform and are already provided to the end users. Therefore, novel building blocks taking advantage of these additional functionalities will be developed to further enhance the existing selection of building blocks on the SOI platform.

\section{PCM/III-V augmented SOI building blocks}
\begin{figure}[h]
  \centering
  \includegraphics[width=0.95\linewidth]{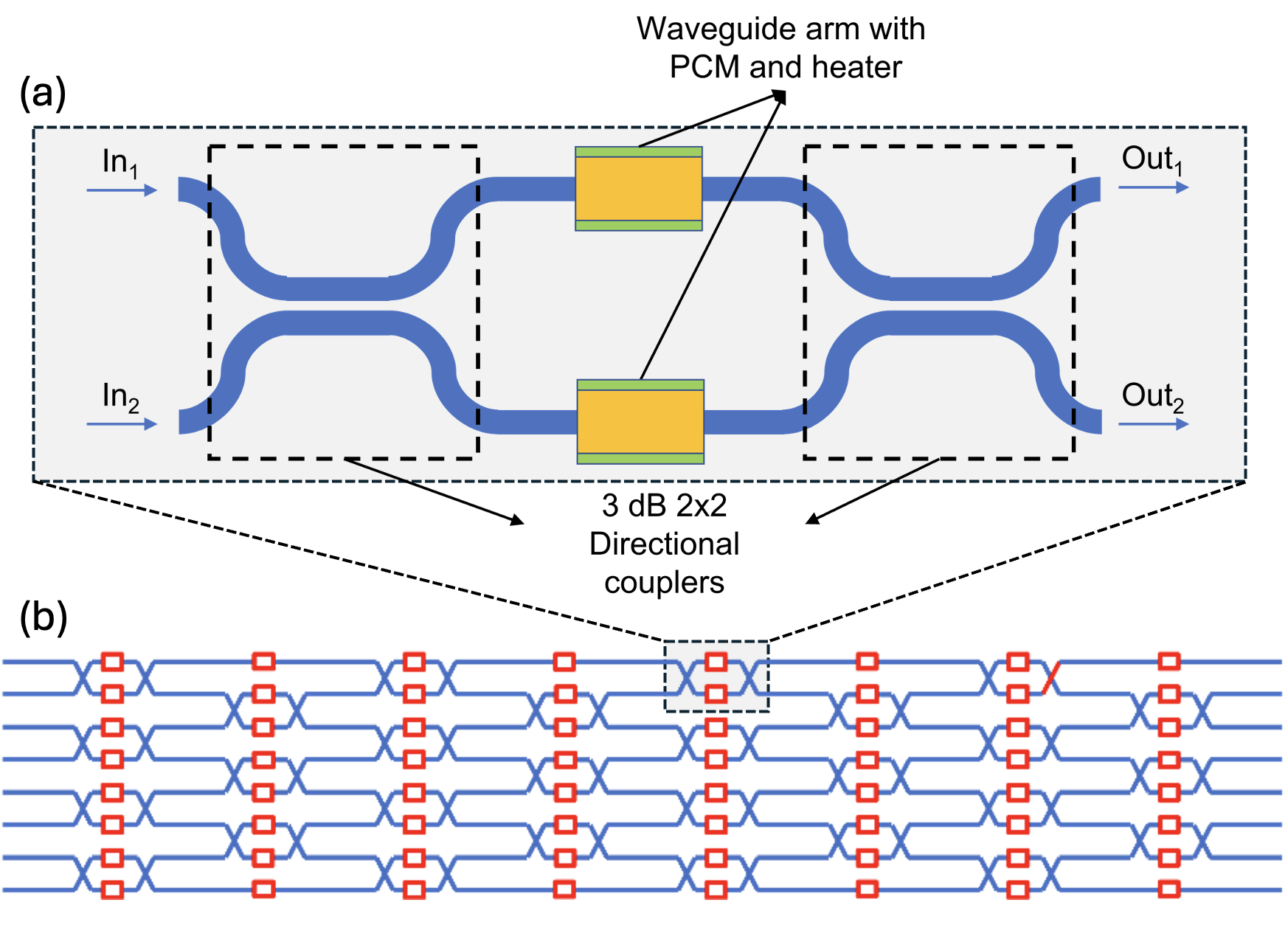}
  \caption{(a) MZI with the PCM-augmented (in green) non-volatile optical phase-shifters with heaters on top (in yellow) for programmability. (b) An example of an MZI mesh architecture (here implementing $8 \times 8$ matrix) dedicated to accelerating matrix-vector multiplication operations via in-memory optical computing.}
  \label{fig:MZI}
\end{figure}

One of the key building blocks for broadband neuromorphic photonic architectures is the Mach-Zenhder interferometer (MZI, shown in Figure 2(a)). An individual MZI consists of couplers and phase-shifters (PS) and represents a $SU(2)$ transformation.
Typically in SOI, a specific phase-shift is induced through the thermo-optic effect via an adjacent heater, and continuously consumes electrical power. Given that this phase-shift remains constant for a set weight matrix (that is, during inference), a non-volatile approach would be ideal to remove this constant energy consumption \cite{miscuglio_artificial_nodate}. Furthermore, besides the non-volatile nature of optical phase shift or attenuation effects, the devices should be compact with minimized optical loss to enable deep arrangements of MZIs. One of the goals of NEUROPULS is the development of low-loss, compact, and reconfigurable multilevel PCM-based MZIs with heaters above PCM patches and waveguides (see Figure~\ref{fig:MZI}(a)). Various approaches are currently being investigated to benefit from the properties of PCMs such as GeSe and GSST that present a large figure of merit (FOM) given by $\delta n/ \delta k$, where $\delta n$ and $\delta k$ are the refractive index contrasts for the real and imaginary part, respectively, around the standard telecom wavelength of 1550 nm \cite{soref_electro-optical_2015,dory_gesbssete_2020}.

Furthermore, Q-switched III-V on-chip lasers are explored as chipscale excitable spiking sources. Pioneering works in this direction have already been carried out; however, spikes were generated off-chip, unlike the approach that we will be pursuing in NEUROPULS \cite{feldmann_all-optical_2019}. By leveraging the ultrafast response (sub-ns) and accumulation behavior of PCM-based devices to optical pulses, the viability of photonic spiking neural networks (SNN) and bio-inspired learning rules such as spike-timing dependent plasticity (STDP) will be investigated.

\section{Computing architectures}
The main focus of the photonic neuromorphic architecture is to realize optical in-memory acceleration of linear algebra operations that underpin a majority of current deep learning models. In particular, we focus on realizing a photonic matrix-vector multiplication (MVM) engine to enable a generalized matrix-matrix (GeMM) core. These architectures are based on meshes of programmable integrated MZIs that operate as multiport interferometers with a degree of matrix expressivity (universality) determined by component arrangement.
Within the NEUROPULS project, various mesh architectures of MZIs (or, more generally, couplers and phase-shifters) are investigated and evaluated. These include previously proposed mesh architectures such as the Clements \cite{clements_optimal_2016} architecture with compacted interferometers \cite{Bell2021_APLPhot} (shown in Figure 2(b)) or the Fldzhyan \cite{Fldzhyan2020_OL} architecture with parallel PS blocks \cite{Bell2021_APLPhot}, as well as newly proposed multiport interferometer architectures. In these, input vectors are encoded into amplitude/phase of individual inputs (typically using high-speed Mach Zehnder modulators), and the multiplication (weighting) matrix is encoded in the state of the programmable PS blocks. Generalization to GeMM operations can be realized through separating of the input matrix into rows, and processing those either via time-division multiplexing or through encoding into multiple dense wavelength division multiplexed (DWDM) channels that can be processed in parallel in a single multiport interferometer without incurring additional resource costs.

\section{Simulation platform}
In the ever-evolving landscape of computing systems, the integration of neuromorphic accelerators and hardware security primitives on a consolidated simulation platform is crucial. Both electronic and photonic domain-specific accelerators (DSAs) continue to be active areas of research and development as the demand for specialized and efficient computing solutions grows across various industries

\begin{figure}[h]
  \centering
  \includegraphics[width=\linewidth]{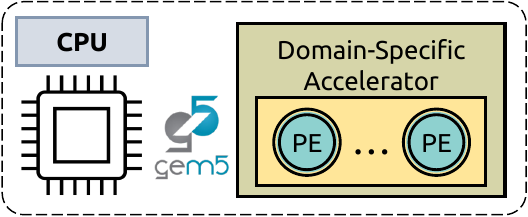}
  \caption{gem5-based system architecture modeling and simulation infrastructure overview.}
  \label{fig:sim}
\end{figure}

In the NEUROPULS project, we will: (a) create efficient full system simulation tools on top of the gem5 simulator~\cite{10.1145/2024716.2024718} to model and evaluate complete computing systems with neuromorphic accelerators and security primitives, as shown in Figure~\ref{fig:sim}; (b) explore the diverse design space of heterogeneous computing systems employing photonic neuromorphic accelerators and hardware primitives; and (c) facilitate detailed system-level evaluation of both software and hardware, with a specific emphasis on the security properties of the computing platform. 

NEUROPULS simulator is based on gem5, an open-source, system-level computer architecture simulator that provides a flexible and modular framework for modeling and simulating various aspects of computer systems. It supports cycle-level simulation of a wide range of computer architectures, including x86, Arm, RISC-V, and others, making it a versatile platform at the system level (including the microarchitecture, architecture, operating systems, and application layers of the computing stack). Given the tremendous growth of the RISC-V ecosystem in the past few years, we have taken gem5-SALAM, which uses an advanced dynamic graph execution engine based on LLVM \cite{9251937} and only supports Arm ISA processor cores (the CPU part in Figure~\ref{fig:sim}), and ported it to support the RISC-V ISA and system configuration. We introduce gem5-MARVEL~\cite{hpca2024-chatzop}, which is based on LLVM IR (Intermediate Representation) to model DSAs using C descriptions of their functionality.
The gem5-based simulation infrastructure comprises two core components: the Compute Unit and the Communications Interface. The Compute Unit represents the custom accelerator's datapath, while the Communications Interface facilitates memory access, control, and synchronization through memory access ports, Memory-Mapped Registers (MMRs), and interrupt lines. The memory access ports allow parallel access to different memory types, such as scratchpad memories (SPMs) and register banks (these two types of memories occupy the largest part of the area of many accelerators). MMRs consist of configurable status, control, and data registers, allowing low-level device configuration and facilitating communication between the accelerator and the host, as well as between multiple accelerators (i.e., processing elements - PEs) in a cluster (as shown in the right-most side of Figure~\ref{fig:sim}). By treating the accelerator as a memory-mapped device, the host can utilize the provided interrupt signals for synchronization without the need for constant polling.  Additionally, the gem5-based infrastructure includes Direct Memory Access (DMA) devices and custom memories that can be seamlessly integrated into accelerator designs, enhancing its versatility. gem5-MARVEL is also a fault injection framework that operates at the microarchitecture level and supports transient and permanent fault injections to all hardware structures of the CPU and for the three prevailing ISAs (Arm, x86, RISC-V). The fault injection feature was implemented in the simulation framework to support the reliability aspect of the NEUROPULS project.

\section{Conclusions}

Integrated photonics represent one of the technological platforms with potential to empower modern heterogeneous computing systems by enabling high bandwidth and improved energy efficiency during data movement and computing. While silicon photonics benefits from compatibility with CMOS technology, further augmentation with additional material platforms is required to unlock the full scale of needed chipscale optical building blocks. In the NEUROPULS project, we have augmented a CMOS-compatible SOI photonic platform with III-V semiconductor technology (enabling on-chip active optical devices), and chalcogenide-based PCMs (to realize non-volatile optical modulators). Using this augmented platform, we propose and evaluate a system-level implementation of a neuromorphic accelerator based on an in-memory photonic MVM accelerator core. Various MZI mesh architectures are evaluated for the MVM core, including their performance, matrix expressivity and robustness.
Furthermore, we have developed a novel gem5-based simulation framework with RISC-V ISA support to allow for extensive performance evaluation and benchmarking of the complete photonic-enabled accelerator interfaced with controllers and processors. We believe this comprehensive system-level implementation is key to realize practical, photonic neuromorphic accelerator.

\begin{acks}
This work has received funding from the European Union’s Horizon Europe research and innovation program under grant agreement No. 101070238. 
Views and opinions expressed are however those of the author(s) only and do not necessarily reflect those of the European Union. Neither the European Union nor the granting authority can be held responsible for them.
\end{acks}



\bibliographystyle{ACM-Reference-Format2}

\bibliography{refs_mhejda,refs-simulation,materials,neuromorphic,photonic_devices_systems,computing_paradigms}


\begin{thebibliography}{23}


\ifx \showCODEN    \undefined \def \showCODEN     #1{\unskip}     \fi
\ifx \showDOI      \undefined \def \showDOI       #1{#1}\fi
\ifx \showISBNx    \undefined \def \showISBNx     #1{\unskip}     \fi
\ifx \showISBNxiii \undefined \def \showISBNxiii  #1{\unskip}     \fi
\ifx \showISSN     \undefined \def \showISSN      #1{\unskip}     \fi
\ifx \showLCCN     \undefined \def \showLCCN      #1{\unskip}     \fi
\ifx \shownote     \undefined \def \shownote      #1{#1}          \fi
\ifx \showarticletitle \undefined \def \showarticletitle #1{#1}   \fi
\ifx \showURL      \undefined \def \showURL       {\relax}        \fi
\providecommand\bibfield[2]{#2}
\providecommand\bibinfo[2]{#2}
\providecommand\natexlab[1]{#1}
\providecommand\showeprint[2][]{arXiv:#2}

\bibitem[Bell and Walmsley(2021)]%
        {Bell2021_APLPhot}
\bibfield{author}{\bibinfo{person}{B.~A. Bell} {and} \bibinfo{person}{I.~A. Walmsley}.} \bibinfo{year}{2021}\natexlab{}.
\newblock \showarticletitle{Further Compactifying Linear Optical Unitaries}.
\newblock \bibinfo{journal}{\emph{APL Photonics}} \bibinfo{volume}{6}, \bibinfo{number}{7} (\bibinfo{date}{July} \bibinfo{year}{2021}), \bibinfo{pages}{070804}.
\newblock
\showISSN{2378-0967}
\urldef\tempurl%
\url{https://doi.org/10.1063/5.0053421}
\showDOI{\tempurl}


\bibitem[Binkert et~al\mbox{.}(2011)]%
        {10.1145/2024716.2024718}
\bibfield{author}{\bibinfo{person}{Nathan Binkert}, \bibinfo{person}{Bradford Beckmann}, \bibinfo{person}{Gabriel Black}, \bibinfo{person}{Steven~K. Reinhardt}, \bibinfo{person}{Ali Saidi}, {et~al\mbox{.}}} \bibinfo{year}{2011}\natexlab{}.
\newblock \showarticletitle{The Gem5 Simulator}.
\newblock \bibinfo{journal}{\emph{SIGARCH Comput. Archit. News}} \bibinfo{volume}{39}, \bibinfo{number}{2} (\bibinfo{date}{aug} \bibinfo{year}{2011}), \bibinfo{pages}{1–7}.
\newblock
\showISSN{0163-5964}
\urldef\tempurl%
\url{https://doi.org/10.1145/2024716.2024718}
\showDOI{\tempurl}


\bibitem[Cao et~al\mbox{.}(2020)]%
        {cao_overview_2020}
\bibfield{author}{\bibinfo{person}{Keyan Cao}, \bibinfo{person}{Yefan Liu}, \bibinfo{person}{Gongjie Meng}, {and} \bibinfo{person}{Qimeng Sun}.} \bibinfo{year}{2020}\natexlab{}.
\newblock \showarticletitle{An Overview on Edge Computing Research}.
\newblock \bibinfo{journal}{\emph{IEEE Access}}  \bibinfo{volume}{8} (\bibinfo{year}{2020}), \bibinfo{pages}{85714--85728}.
\newblock
\urldef\tempurl%
\url{https://doi.org/10.1109/ACCESS.2020.2991734}
\showDOI{\tempurl}


\bibitem[Chatzopoulos et~al\mbox{.}(2024)]%
        {hpca2024-chatzop}
\bibfield{author}{\bibinfo{person}{Odysseas Chatzopoulos}, \bibinfo{person}{George Papadimitriou}, \bibinfo{person}{Vasileios Karakostas}, {and} \bibinfo{person}{Dimitris Gizopoulos}.} \bibinfo{year}{2024}\natexlab{}.
\newblock \showarticletitle{gem5-MARVEL: Microarchitecture-Level Resilience Analysis of Heterogeneous SoC Architectures}. In \bibinfo{booktitle}{\emph{IEEE International Symposium on High-Performance Computer Architecture (HPCA 2024)}}. \bibinfo{pages}{543--559}.
\newblock
\urldef\tempurl%
\url{https://doi.org/10.1109/HPCA57654.2024.00047}
\showDOI{\tempurl}


\bibitem[Clements et~al\mbox{.}(2016)]%
        {clements_optimal_2016}
\bibfield{author}{\bibinfo{person}{William~R. Clements}, \bibinfo{person}{Peter~C. Humphreys}, \bibinfo{person}{Benjamin~J. Metcalf}, \bibinfo{person}{W.~Steven Kolthammer}, {and} \bibinfo{person}{Ian~A. Walsmley}.} \bibinfo{year}{2016}\natexlab{}.
\newblock \showarticletitle{Optimal design for universal multiport interferometers}.
\newblock \bibinfo{journal}{\emph{Optica}} \bibinfo{volume}{3}, \bibinfo{number}{12} (\bibinfo{date}{Dec.} \bibinfo{year}{2016}), \bibinfo{pages}{1460}.
\newblock
\showISSN{2334-2536}
\urldef\tempurl%
\url{https://doi.org/10.1364/OPTICA.3.001460}
\showDOI{\tempurl}


\bibitem[Davies et~al\mbox{.}(2021)]%
        {davies_advancing_2021}
\bibfield{author}{\bibinfo{person}{Mike Davies}, \bibinfo{person}{Andreas Wild}, \bibinfo{person}{Garrick Orchard}, \bibinfo{person}{Yulia Sandamirskaya}, \bibinfo{person}{Gabriel A.~Fonseca Guerra}, {et~al\mbox{.}}} \bibinfo{year}{2021}\natexlab{}.
\newblock \showarticletitle{Advancing {Neuromorphic} {Computing} {With} {Loihi}: {A} {Survey} of {Results} and {Outlook}}.
\newblock \bibinfo{journal}{\emph{Proc. IEEE}} \bibinfo{volume}{109}, \bibinfo{number}{5} (\bibinfo{date}{May} \bibinfo{year}{2021}), \bibinfo{pages}{911--934}.
\newblock
\showISSN{1558-2256}
\urldef\tempurl%
\url{https://doi.org/10.1109/JPROC.2021.3067593}
\showDOI{\tempurl}
\newblock
\shownote{Conference Name: Proceedings of the IEEE}.


\bibitem[Dory et~al\mbox{.}(2020)]%
        {dory_gesbssete_2020}
\bibfield{author}{\bibinfo{person}{J.-B. Dory}, \bibinfo{person}{C. Castro-Chavarria}, \bibinfo{person}{A. Verdy}, \bibinfo{person}{J.-B. Jager}, \bibinfo{person}{M. Bernard}, {et~al\mbox{.}}} \bibinfo{year}{2020}\natexlab{}.
\newblock \showarticletitle{Ge–{Sb}–{S}–{Se}–{Te} amorphous chalcogenide thin films towards on-chip nonlinear photonic devices}.
\newblock \bibinfo{journal}{\emph{Sci Rep}} \bibinfo{volume}{10}, \bibinfo{number}{1} (\bibinfo{date}{July} \bibinfo{year}{2020}), \bibinfo{pages}{11894}.
\newblock
\showISSN{2045-2322}
\urldef\tempurl%
\url{https://doi.org/10.1038/s41598-020-67377-9}
\showDOI{\tempurl}


\bibitem[Feldmann et~al\mbox{.}(2021)]%
        {feldmann_parallel_2021}
\bibfield{author}{\bibinfo{person}{J. Feldmann}, \bibinfo{person}{N. Youngblood}, \bibinfo{person}{M. Karpov}, \bibinfo{person}{H. Gehring}, \bibinfo{person}{X. Li}, {et~al\mbox{.}}} \bibinfo{year}{2021}\natexlab{}.
\newblock \showarticletitle{Parallel convolutional processing using an integrated photonic tensor core}.
\newblock \bibinfo{journal}{\emph{Nature}} \bibinfo{volume}{589}, \bibinfo{number}{7840} (\bibinfo{date}{Jan.} \bibinfo{year}{2021}), \bibinfo{pages}{52--58}.
\newblock
\showISSN{0028-0836, 1476-4687}
\urldef\tempurl%
\url{https://doi.org/10.1038/s41586-020-03070-1}
\showDOI{\tempurl}


\bibitem[Feldmann et~al\mbox{.}(2019)]%
        {feldmann_all-optical_2019}
\bibfield{author}{\bibinfo{person}{J. Feldmann}, \bibinfo{person}{N. Youngblood}, \bibinfo{person}{C.~D. Wright}, \bibinfo{person}{H. Bhaskaran}, {and} \bibinfo{person}{W.~H.~P. Pernice}.} \bibinfo{year}{2019}\natexlab{}.
\newblock \showarticletitle{All-optical spiking neurosynaptic networks with self-learning capabilities}.
\newblock \bibinfo{journal}{\emph{Nature}} \bibinfo{volume}{569}, \bibinfo{number}{7755} (\bibinfo{date}{May} \bibinfo{year}{2019}), \bibinfo{pages}{208--214}.
\newblock
\showISSN{0028-0836, 1476-4687}
\urldef\tempurl%
\url{https://doi.org/10.1038/s41586-019-1157-8}
\showDOI{\tempurl}


\bibitem[Fldzhyan et~al\mbox{.}(2020)]%
        {Fldzhyan2020_OL}
\bibfield{author}{\bibinfo{person}{S.~A. Fldzhyan}, \bibinfo{person}{M.~Yu Saygin}, {and} \bibinfo{person}{S.~P. Kulik}.} \bibinfo{year}{2020}\natexlab{}.
\newblock \showarticletitle{Optimal Design of Error-Tolerant Reprogrammable Multiport Interferometers}.
\newblock \bibinfo{journal}{\emph{Optics Letters}} \bibinfo{volume}{45}, \bibinfo{number}{9} (\bibinfo{date}{May} \bibinfo{year}{2020}), \bibinfo{pages}{2632--2635}.
\newblock
\showISSN{1539-4794}
\urldef\tempurl%
\url{https://doi.org/10.1364/OL.385433}
\showDOI{\tempurl}


\bibitem[Frenkel et~al\mbox{.}(2023)]%
        {Frenkel2023_PotI}
\bibfield{author}{\bibinfo{person}{Charlotte Frenkel}, \bibinfo{person}{David Bol}, {and} \bibinfo{person}{Giacomo Indiveri}.} \bibinfo{year}{2023}\natexlab{}.
\newblock \showarticletitle{Bottom-{{Up}} and {{Top-Down Approaches}} for the {{Design}} of {{Neuromorphic Processing Systems}}: {{Tradeoffs}} and {{Synergies Between Natural}} and {{Artificial Intelligence}}}.
\newblock \bibinfo{journal}{\emph{Proc. IEEE}} \bibinfo{volume}{111}, \bibinfo{number}{6} (\bibinfo{date}{June} \bibinfo{year}{2023}), \bibinfo{pages}{623--652}.
\newblock
\showISSN{1558-2256}
\urldef\tempurl%
\url{https://doi.org/10.1109/JPROC.2023.3273520}
\showDOI{\tempurl}


\bibitem[Grollier et~al\mbox{.}(2020)]%
        {grollier_neuromorphic_2020}
\bibfield{author}{\bibinfo{person}{J. Grollier}, \bibinfo{person}{D. Querlioz}, \bibinfo{person}{K.~Y. Camsari}, \bibinfo{person}{K. Everschor-Sitte}, \bibinfo{person}{S. Fukami}, {et~al\mbox{.}}} \bibinfo{year}{2020}\natexlab{}.
\newblock \showarticletitle{Neuromorphic spintronics}.
\newblock \bibinfo{journal}{\emph{Nat Electron}} \bibinfo{volume}{3}, \bibinfo{number}{7} (\bibinfo{date}{July} \bibinfo{year}{2020}), \bibinfo{pages}{360--370}.
\newblock
\showISSN{2520-1131}
\urldef\tempurl%
\url{https://doi.org/10.1038/s41928-019-0360-9}
\showDOI{\tempurl}
\newblock
\shownote{Publisher: Nature Publishing Group}.


\bibitem[Höppner et~al\mbox{.}(2022)]%
        {hoppner_spinnaker_2022}
\bibfield{author}{\bibinfo{person}{Sebastian Höppner}, \bibinfo{person}{Yexin Yan}, \bibinfo{person}{Andreas Dixius}, \bibinfo{person}{Stefan Scholze}, \bibinfo{person}{Johannes Partzsch}, {et~al\mbox{.}}} \bibinfo{year}{2022}\natexlab{}.
\newblock \bibinfo{title}{The {SpiNNaker} 2 {Processing} {Element} {Architecture} for {Hybrid} {Digital} {Neuromorphic} {Computing}}.
\newblock
\newblock
\urldef\tempurl%
\url{http://arxiv.org/abs/2103.08392}
\showURL{%
\tempurl}


\bibitem[Miscuglio et~al\mbox{.}({[n.\,d.]})]%
        {miscuglio_artificial_nodate}
\bibfield{author}{\bibinfo{person}{Mario Miscuglio}, \bibinfo{person}{Jiawei Meng}, \bibinfo{person}{Omer Yesiliurt}, \bibinfo{person}{Yifei Zhang}, \bibinfo{person}{Ludmila~J Prokopeva}, {et~al\mbox{.}}} \bibinfo{year}{[n.\,d.]}\natexlab{}.
\newblock \showarticletitle{Artificial {Synapse} with {Mnemonic} {Functionality} using {GSST}-based {Photonic} {Integrated} {Memory}}.
\newblock  (\bibinfo{year}{[n.\,d.]}), \bibinfo{pages}{8}.
\newblock


\bibitem[Pavanello et~al\mbox{.}(2023)]%
        {pavanello_neuropuls_2023}
\bibfield{author}{\bibinfo{person}{Fabio Pavanello}, \bibinfo{person}{Cedric Marchand}, \bibinfo{person}{Ian O’Connor}, \bibinfo{person}{Regis Orobtchouk}, \bibinfo{person}{Fabien Mandorlo}, {et~al\mbox{.}}} \bibinfo{year}{2023}\natexlab{}.
\newblock \showarticletitle{{NEUROPULS}: {NEUROmorphic} energy-efficient secure accelerators based on {Phase} change materials {aUgmented} {siLicon} {photonicS}}.
\newblock \bibinfo{journal}{\emph{2023 IEEE European Test Symposium (ETS)}}.
\newblock
\urldef\tempurl%
\url{https://doi.org/10.1109/ETS56758.2023.10173974}
\showDOI{\tempurl}


\bibitem[Rahim et~al\mbox{.}(2018)]%
        {rahim_open-access_2018}
\bibfield{author}{\bibinfo{person}{Abdul Rahim}, \bibinfo{person}{Thijs Spuesens}, \bibinfo{person}{Roel Baets}, {and} \bibinfo{person}{Wim Bogaerts}.} \bibinfo{year}{2018}\natexlab{}.
\newblock \showarticletitle{Open-{Access} {Silicon} {Photonics}: {Current} {Status} and {Emerging} {Initiatives}}.
\newblock \bibinfo{journal}{\emph{Proc. IEEE}} \bibinfo{volume}{106}, \bibinfo{number}{12} (\bibinfo{date}{Dec.} \bibinfo{year}{2018}), \bibinfo{pages}{2313--2330}.
\newblock
\showISSN{1558-2256}
\urldef\tempurl%
\url{https://doi.org/10.1109/JPROC.2018.2878686}
\showDOI{\tempurl}
\newblock
\shownote{Conference Name: Proceedings of the IEEE}.


\bibitem[Rogers et~al\mbox{.}(2020)]%
        {9251937}
\bibfield{author}{\bibinfo{person}{Samuel Rogers}, \bibinfo{person}{Joshua Slycord}, \bibinfo{person}{Mohammadreza Baharani}, {and} \bibinfo{person}{Hamed Tabkhi}.} \bibinfo{year}{2020}\natexlab{}.
\newblock \showarticletitle{gem5-SALAM: A System Architecture for LLVM-based Accelerator Modeling}. In \bibinfo{booktitle}{\emph{2020 53rd Annual IEEE/ACM International Symposium on Microarchitecture (MICRO)}}. \bibinfo{pages}{471--482}.
\newblock
\urldef\tempurl%
\url{https://doi.org/10.1109/MICRO50266.2020.00047}
\showDOI{\tempurl}


\bibitem[Samek et~al\mbox{.}(2021)]%
        {samek_explaining_2021}
\bibfield{author}{\bibinfo{person}{Wojciech Samek}, \bibinfo{person}{Gregoire Montavon}, \bibinfo{person}{Sebastian Lapuschkin}, \bibinfo{person}{Christopher~J. Anders}, {and} \bibinfo{person}{Klaus-Robert Muller}.} \bibinfo{year}{2021}\natexlab{}.
\newblock \showarticletitle{Explaining {Deep} {Neural} {Networks} and {Beyond}: {A} {Review} of {Methods} and {Applications}}.
\newblock \bibinfo{journal}{\emph{Proc. IEEE}} \bibinfo{volume}{109}, \bibinfo{number}{3} (\bibinfo{date}{March} \bibinfo{year}{2021}), \bibinfo{pages}{247--278}.
\newblock
\showISSN{0018-9219, 1558-2256}
\urldef\tempurl%
\url{https://doi.org/10.1109/JPROC.2021.3060483}
\showDOI{\tempurl}


\bibitem[Satyanarayanan(2017)]%
        {satyanarayanan_emergence_2017}
\bibfield{author}{\bibinfo{person}{Mahadev Satyanarayanan}.} \bibinfo{year}{2017}\natexlab{}.
\newblock \showarticletitle{The {Emergence} of {Edge} {Computing}}.
\newblock \bibinfo{journal}{\emph{Computer}} \bibinfo{volume}{50}, \bibinfo{number}{1} (\bibinfo{date}{Jan.} \bibinfo{year}{2017}), \bibinfo{pages}{30--39}.
\newblock
\showISSN{0018-9162}
\urldef\tempurl%
\url{https://doi.org/10.1109/MC.2017.9}
\showDOI{\tempurl}


\bibitem[Shastri et~al\mbox{.}(2021)]%
        {Shastri2021_NatPhot}
\bibfield{author}{\bibinfo{person}{Bhavin~J. Shastri}, \bibinfo{person}{Alexander~N. Tait}, \bibinfo{person}{Thomas {Ferreira de Lima}}, \bibinfo{person}{Wolfram H.~P. Pernice}, \bibinfo{person}{Harish Bhaskaran}, {et~al\mbox{.}}} \bibinfo{year}{2021}\natexlab{}.
\newblock \showarticletitle{Photonics for Artificial Intelligence and Neuromorphic Computing}.
\newblock \bibinfo{journal}{\emph{Nature Photonics}} \bibinfo{volume}{15}, \bibinfo{number}{2} (\bibinfo{date}{Feb.} \bibinfo{year}{2021}), \bibinfo{pages}{102--114}.
\newblock
\showISSN{1749-4885, 1749-4893}
\urldef\tempurl%
\url{https://doi.org/10.1038/s41566-020-00754-y}
\showDOI{\tempurl}


\bibitem[Soref et~al\mbox{.}(2015)]%
        {soref_electro-optical_2015}
\bibfield{author}{\bibinfo{person}{Richard Soref}, \bibinfo{person}{Joshua Hendrickson}, \bibinfo{person}{Haibo Liang}, \bibinfo{person}{Arka Majumdar}, \bibinfo{person}{Jianwei Mu}, {et~al\mbox{.}}} \bibinfo{year}{2015}\natexlab{}.
\newblock \showarticletitle{Electro-optical switching at 1550 nm using a two-state {GeSe} phase-change layer}.
\newblock \bibinfo{journal}{\emph{Opt. Express, OE}} \bibinfo{volume}{23}, \bibinfo{number}{2} (\bibinfo{date}{Jan.} \bibinfo{year}{2015}), \bibinfo{pages}{1536--1546}.
\newblock
\showISSN{1094-4087}
\urldef\tempurl%
\url{https://doi.org/10.1364/OE.23.001536}
\showDOI{\tempurl}
\newblock
\shownote{Publisher: Optica Publishing Group}.


\bibitem[Tait et~al\mbox{.}(2019)]%
        {tait_silicon_2019}
\bibfield{author}{\bibinfo{person}{Alexander~N. Tait}, \bibinfo{person}{Thomas Ferreira~de Lima}, \bibinfo{person}{Mitchell~A. Nahmias}, \bibinfo{person}{Heidi~B. Miller}, \bibinfo{person}{Hsuan-Tung Peng}, {et~al\mbox{.}}} \bibinfo{year}{2019}\natexlab{}.
\newblock \showarticletitle{Silicon {Photonic} {Modulator} {Neuron}}.
\newblock \bibinfo{journal}{\emph{Phys. Rev. Appl.}} \bibinfo{volume}{11}, \bibinfo{number}{6} (\bibinfo{date}{June} \bibinfo{year}{2019}), \bibinfo{pages}{064043}.
\newblock
\urldef\tempurl%
\url{https://doi.org/10.1103/PhysRevApplied.11.064043}
\showDOI{\tempurl}
\newblock
\shownote{Publisher: American Physical Society}.


\bibitem[Zhou et~al\mbox{.}(2023)]%
        {zhou_-memory_2023}
\bibfield{author}{\bibinfo{person}{Wen Zhou}, \bibinfo{person}{Bowei Dong}, \bibinfo{person}{Nikolaos Farmakidis}, \bibinfo{person}{Xuan Li}, \bibinfo{person}{Nathan Youngblood}, {et~al\mbox{.}}} \bibinfo{year}{2023}\natexlab{}.
\newblock \showarticletitle{In-memory photonic dot-product engine with electrically programmable weight banks}.
\newblock \bibinfo{journal}{\emph{Nat Commun}} \bibinfo{volume}{14}, \bibinfo{number}{1} (\bibinfo{date}{May} \bibinfo{year}{2023}), \bibinfo{pages}{2887}.
\newblock
\showISSN{2041-1723}
\urldef\tempurl%
\url{https://doi.org/10.1038/s41467-023-38473-x}
\showDOI{\tempurl}


\end{thebibliography}

\end{document}